\documentclass[conference]{IEEEtran}
\IEEEoverridecommandlockouts
\usepackage{tabularx,ragged2e,booktabs}
\usepackage{float}
\usepackage{graphicx}
\usepackage{xcolor}
\usepackage{graphics}
\usepackage{mathtools}
\usepackage{physics}
\usepackage{color}
\usepackage{amsfonts}
\usepackage{xcolor}
\usepackage{tikz}
\usepackage[hyphens]{url}
\usepackage{graphicx}
\usepackage{placeins}
\usepackage{amsmath}
\usepackage{amssymb}
\usepackage{pdfpages}
\usepackage{todonotes}
\usepackage{threeparttable}
\usepackage{dcolumn}
\usepackage{wrapfig}
\usepackage{glossaries}
\usepackage[symbol]{footmisc}
\usepackage{subcaption}
\usepackage{soul}
\usepackage{float}
\usepackage{psfrag}
\usepackage{enumitem}
\definecolor{gray1}{gray}{0.7}
\definecolor{gray2}{gray}{0.9}

\begin{document}

\title{Metamodel Based Forward and Inverse Design for Passive Vibration Suppression
\thanks{Copyright\copyright 2020 ASME. Personal use of this material is permitted. Permission from ASME must be obtained for all other uses, in any current or future media, including reprinting/republishing this material for advertising or promotional purposes, creating new collective works, for resale or redistribution to servers or lists, or reuse of any copyrighted component of this work in other works.}
}

\author{\IEEEauthorblockN{Amir Behjat\IEEEauthorrefmark{1}, 
Manaswin Oddiraju\IEEEauthorrefmark{2}, 
Mohammad Ali Attarzadeh\IEEEauthorrefmark{3}, 
Mostafa Nouh\IEEEauthorrefmark{4}, 
Souma Chowdhury\IEEEauthorrefmark{5}}
\IEEEauthorblockA{\textit{Department of Mechanical \& Aerospace Engineering} \\
\textit{University at Buffalo}\\
Buffalo, NY, 14260\\
Email: \IEEEauthorrefmark{1}amirbehj@buffalo.edu,
\IEEEauthorrefmark{2}moddiraj@buffalo.edu,
\IEEEauthorrefmark{3}mattarza@buffalo.edu,
\IEEEauthorrefmark{4}mnouh@buffalo.edu,
\IEEEauthorrefmark{5}soumacho@buffalo.edu}}

\maketitle  
\begin{abstract}

Aperiodic metamaterials represent a class of structural systems that are composed of different building blocks (cells), instead of a self-repeating chain of the same unit cells. Optimizing aperiodic cellular structural systems thus presents high-dimensional design problems, that become intractable to solve using purely high-fidelity structural analysis coupled with optimization. Specialized analytical modeling along with metamodel based optimization can provide a more tractable alternative to designing such aperiodic metamaterials. To explore this concept, this paper presents an initial design automation framework applied to a case study representative of a simple 1D metamaterial system. The case under consideration is a drill string, where vibration suppression is of utmost importance. The drill string comprises a set of nonuniform rings attached to the outer surface of a longitudinal rod. As such, the resultant system can now be perceived as an aperiodic 1D metamaterial with each ring/gap representing a cell. Despite being a 1D system, the simultaneous consideration of multiple degrees of freedom (associated with torsional, axial, and lateral motions) poses significant computational challenges. To deal with these challenges, a transfer matrix method (TMM) is employed to analytically determine the frequency response of the drill string. However, due to the minute scale cost of the TMM method, the optimization remains computationally burdensome. This latter challenge is addressed by training a suite of neural networks on a set of TMM samples, with each network providing the response w.r.t. a specific frequency. Optimization is then performed to minimize mass subject to constraints on the gap between consecutive resonance peaks in one case, and minimizing this gap in the second case. Crucial improvements are accomplished over the initial baselines in both cases. Further novel contributions occur through the development of an inverse modeling approach
that can learn optimal inverse designs with minimum mass and a desirable non-resonant frequency range, which partially mimics band gap behavior in perfectly periodic dispersive structures. To this end, we introduce the use of an emerging modeling formalism called invertible neural nets. 
Our study indicates that the inverse model is able to generate constraint satisfying designs with slightly higher mass.

\end{abstract}

\section{Introduction}
 
\subsection{Aperiodic Metamaterial Optimization}
Metamaterials are compound structures that are made from small substructures referred to as ``unit cells" \cite{fok2008acoustic,albabaa2017formation}. Such composition enables them to have customizable physical properties which are otherwise not feasible \cite{fok2008acoustic,chronopoulos2017enhanced}. As such, they have become increasingly popular and, despite manufacturing difficulties \cite{maruyama2010analysis,buckmann2012tailored,yoon2016challenges}, are being widely used in different applications ranging from acoustics \cite{sui2015lightweight,mamaghani2016vibration}, and photonics \cite{liu2008three,gansel2009gold}, to sensing and other end uses \cite{chen2012metamaterials,albabaa2019emergence,nouh2019control}. 


Owing to the periodic nature of traditional metamaterial systems, the common approach utilized in their design predominantly relies on using an individual cell to predict the performance of the overall structure \cite{ge2007ga,nouh2014vibration,nouh2015wave}. 
Contrary to periodic metamaterials, aperiodic designs utilize a more extensive design space opening up more possibilities to achieve a targetted behavior. For instance, Hussein et al. \cite{hussein2006multiobjective} compared the wave dispersion characteristics of a layered metamaterial. By increasing the number of design variables, the performance of the system increased, but the optimization became evidently more challenging. These challenges are expected to further increase by having aperiodic structures such as those presented in \cite{yang2017aperiodic}. In all of these efforts, the optimization was limited by the computational cost. 
There are other approaches to tackle this problem. For instance, Bostanabad et al.  \cite{bostanabad2019globally} used Gaussian Process for big data in a process called Globally Approximate Gaussian Process and used it for metamaterial design. An issue with using Gaussian Process models is the difficulty of inverting the model. While it is possible to use an inverse design based on the GP based surrogate model, it is not necessarily a computationally efficient process, and more importantly a separate optimization is required every time a different end property is desired by the user. Another recent approach uses a Bayesian Network Classifier \cite{matthews2016hierarchical,morris2018design}, which significantly reduces computational effort and provisions model error measures. However, it is not directly amenable to efficient inverse design or on-demand retrieval of metamaterial design given desired properties. 

\textit{Our idea is to use an invertible artificial neural network (ANN) that can generate the inverse design without significant additional computational cost beyond sampling a set of forward optimizations. More specifically, we show that by training a forward ANN based metamodel with certain constraints on its architecture, one can implicitly learn the inverse mapping. To our knowledge, this is the first formal exploration of how to use this nascent modeling paradigm, i.e., provably invertible neural nets, to enable on-demand inverse design.} 
In principle, the method is extendable (in future) to exploit active sampling over optimizations, to offer unprecedented efficiency in inverse metamaterial design or on-demand retrieval of metamaterial topologies. 
In this paper, we specifically use a set of surrogate models to estimate the attenuation response, i.e., substitute the more expensive forward model, which are then used to perform the sample forward optimizations.  

Surrogate based optimization is a class of optimization methods that use surrogate models to decrease computation expense for evaluating the objective function or for constraint violation assessment \cite{han2012surrogate}. Since, physics based models for dynamic analysis of structures tend to be computationally expensive, using surrogate models to represent them is an attractive choice from a computational time savings perspective \cite{Bhosekarreview}.  In order to achieve better accuracy with a smaller number of samples, we used one model for each frequency, which leads to a massive number of models. Using a single model for all outputs could in principle provide more parsimonious representations due to the dependency between outputs \cite{poggio2017and}. However, our current choice of a surrogate per frequency is driven by the observation that, any advantage offered by capturing the dependency between outputs is undermined by the massive output to input ratio (480 outputs for only 30 inputs) that such a model must handle. 
In this study, we successfully train a suite of neural networks that can be used as surrogate models for the optimization of aperiodic structures. 
In order to test our models, we use a passive vibration problem, which is designed inspired by \cite{alsaffar2018band}. In this problem, we use material inserts on a drilling pipe to generate a passive filter for vibration. We use our method in this example by considering the inserts as different unit cells and optimize the drilling pipe to exhibit the required non-resonant frequency range. 


The \textbf{overall objectives} of this paper can be summarized as:
\begin{enumerate}
    \item Develop an efficient forward optimization framework that integrates transfer matrix models and surrogate models to efficiently design 1D metamaterials that minimize mass subject to certain frequency response constraints. 
    \item Investigate a novel invertible neural network approach to learn the forward optimization outcomes in a manner that allows reliable inverse computations for on demand design (given desired properties). 
    \item Demonstrate the effectiveness of the forward and inverse design methods by applying it to design 1D drill strings with multiple modes of vibration.  
\end{enumerate}

\subsection{1D Metamaterial Example: Drill Strings}

Drill strings are usually very long structures composed of thin-walled drill pipes\cite{reviewghasemloonia} used in a wide range of application such as oil exploration and sample collection by planetary rovers. Their inherent slender geometry makes them susceptible to vibrations. 
There are three major types of vibrations that a drill string might face, Longitudinal vibration or Bit-Bounce vibration, which occurs in the axial direction, the Lateral or Whirl vibrations, and Torsional or Stick-Slip vibrations. \cite{alsaffar2018band}. These vibrations can be detrimental to the drill string as well as associated parts including the drill bit, borehole assembly, and well wall \cite{alsaffar2018band,spanos1995modeling,jansen1995active}, which could thus result in major downtimes in drilling  \cite{reviewghasemloonia}. In a study by Alsaffar et al. \cite{alsaffar2018band}, additional mass is added to the drill string to change its dynamic behavior. In this approach, optimally placed periodic inserts in the form of annular rings were attached to the periphery of the drill string. The addition of the inserts, when optimally placed, mitigates high amplitude vibrations, thereby minimizing the possibility of excessive damage to the drill string during operation. Drill strings serve as a suitable example for analyzing the effectiveness of the proposed forward and inverse design methods, since while representing a tractable 1D metamaterial system, it offers substantial complexity in the form of the multiple coupled modes of vibration -- which will help derive general insights from the results obtained from our design case studies. 

The remainder of this paper is structured as follows. In section \ref{sec:trans}, we explain the overall design frameworks, including description of the Transfer Matrix Method for computing the frequency response of the system. In section \ref{sec:prob_def}
 we present the drill string problem formulation. Section \ref{sec:inv_des} explains the inverse design problem and Section \ref{sec:res} is dedicated to the discussion of the design optimization results. In section \ref{sec:cons}, we provide concluding remarks.

\section{Methodology}
\label{sec:trans}


\begin{figure*}[!h]
  \includegraphics[width=\linewidth]{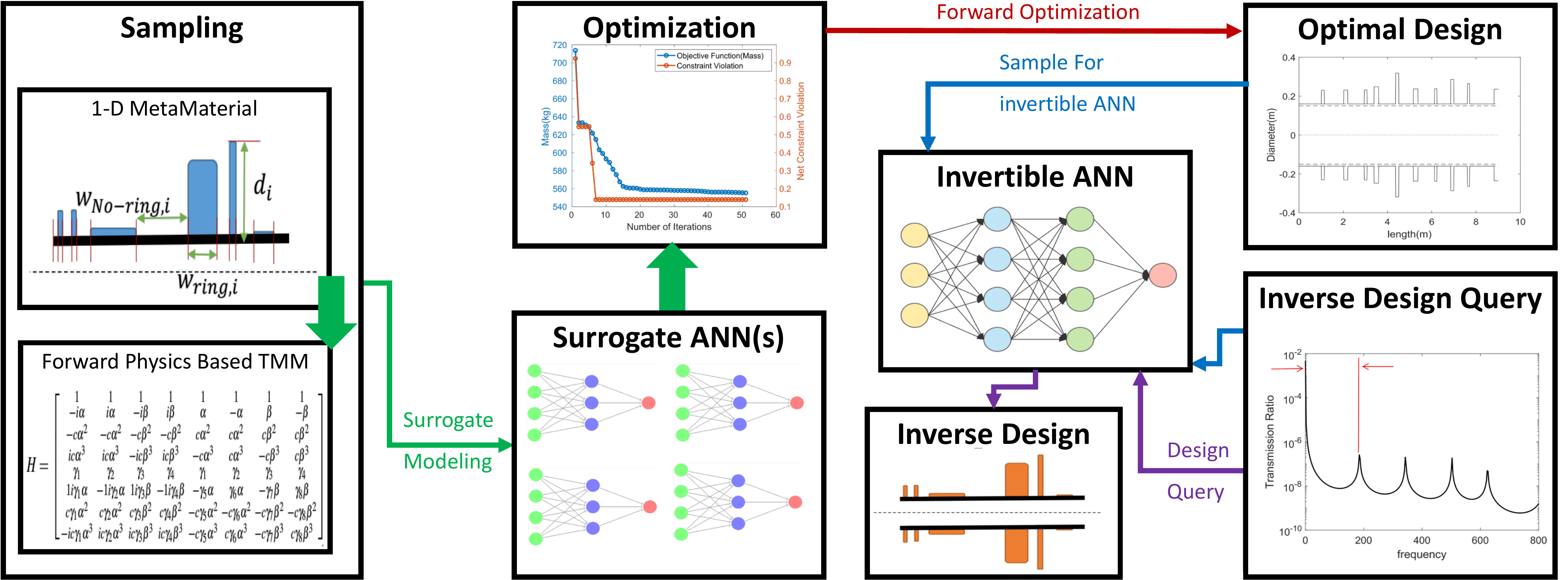}
  \caption{Overall framework}
  \label{fig:framework}
\end{figure*}


\subsection{Forward \& Inverse Design Framework}
Figure \ref{fig:framework} illustrates our overall design automation framework. The goal of this framework is to enable surrogate-based optimization and inverse design to retrieve aperiodic metamaterial configurations given desired passive vibration suppression properties, i.e., frequency response properties. A Drill string is chosen as a case study to analyze the performance of this framework and the benefits of the underlying new inversion concepts. First, a suite of ANNs are trained on samples generated by a physics based forward model, in this case a Transfer matrix method. Then optimization is performed using a Particle Swarm or PSO algorithm \cite{chowdhury2013mixed} to find the design with the maximum non-resonant frequency range. An inverse design mechanism is then developed by performing several optimizations to minimize the mass of the design for several samples of the non-resonant frequency range, with mass minimization acting as a regularizer that seeks to facilitate an inverse that is at least locally unique. An Invertible ANN is then trained on the optimized samples to directly model ``design $\to$ desired-vibration-properties", and implicitly capture the inverse mapping (due to its inherent invertibility). For performance evaluation, we compare the inverse designs (on unseen test samples, i.e., unseen desired property specifications) given by both the invertible ANN model and by solving a constrained forward optimization (where the latter is a conventional approach to inverse design).\\


\subsection{Transfer Matrix Method}
The Transfer Matrix Method (TMM) is an analytical approach with the ability to give an exact solution to the structural dynamics problem. At its core, the TMM tracks the transmission of different forces (or moments as applicable) and deformations (or rotations as applicable) from one segment of a structural system to the adjacent one. As such, to capture the performance of a tapered rod or one of varying cross-section, it must be discretized to different cylindrical sub-bodies. This discretization may lead to a loss of model fidelity as a result of the insufficient number of discrete segments. Another drawback of the TMM method are often some numerical errors associated with calculating the analytical response depending on matrix conditions. Solving the TMM method requires high numerical precision. In this paper, a precision of 100 digits is used in order to avoid errors, which in turn significantly increased the computation time.

The ``Transmission Ratio" refers to the amount of vibration transmitted to one end of a drill pipe when subjected to excitation at the opposing end. This vibration can be modeled as displacement or force based on the boundary conditions. While typically the transmission ratio is below one, during resonance, it increases to a very high value, which can cause significant damage to the structure. In the current study, we obtain the the transmission ratio using the TMM method and optimize the design using surrogate models.
TMM equations for Euler-Bernoulli beam with free-free boundary condition are defined based on Equations \ref{eq:axial1}, \ref{eq:axial2} for Axial Vibrations and equations \ref{eq:torsional1}, \ref{eq:torsional2} for Torsional  vibrations:

\begin{equation}
\begin{aligned}
&T_{i,j} = \begin{bmatrix}
   \cos(\Omega_{i,j}) & \dfrac{\sin(\Omega_{i,j})}{K_{i} \Omega_{i,j}} \\
   -\sin(\Omega_{i,j}) K_{i} \Omega_{i,j} &
   \cos(\Omega_{i,j}) 
   \end{bmatrix}
\end{aligned}
\label{eq:axial1}
\end{equation}

\begin{equation}
    \label{eq:axial2}
    \begin{aligned}
    &C_i=\dfrac{E_i}{\rho_i} ,\  \Omega_{i,j} =\dfrac{2 \pi \omega_{j} w_i}{C_i}, \  K_{i}=\dfrac{E A_{i} }{w_i}
    \end{aligned}
\end{equation}



\begin{equation}
\label{eq:torsional1}
\begin{aligned}
&T_{i,j} = \begin{bmatrix}
   \cos(\Omega_{i,j}) & \dfrac{\sin(\Omega_{i,j})}{K_{i} \Omega_{i,j}} \\
   -\sin(\Omega_{i,j}) K_{i} \Omega_{i,j} &
   \cos(\Omega_{i,j}) 
   \end{bmatrix}
   \end{aligned}
\end{equation}

\begin{equation}
\label{eq:torsional2}
\begin{aligned}
&C_i=\frac{G_i}{\rho_i}
, \Omega_{i,j} =\frac{2 \pi \omega_{j} w_i}{C_i}, \ K_{i}=\dfrac{G_{i} J_{i} }{w_i}, \
\end{aligned}
\end{equation}


\textcolor{black}{While both equation \ref{eq:axial1} and \ref{eq:torsional1} show a decoupled relation to the other DoFs, }
for Lateral Vibrations, we use a set of coupled equations \cite{liu2012wave}. Here we only consider the lateral displacements, and because of the symmetry between two lateral directions, the transmission ratio will be the same for both directions. Therefore, although we need to solve the coupled 8 by 8 problem, we only plot transmission ratio in one direction for lateral vibrations. The equations
below
explain the TMM matrix for this problem:

\begin{flushleft}
\begin{equation}
H = 
\begin{bsmallmatrix}
    1& 1&   1& 1& 1& 1&1&1\\
   -i\alpha& i\alpha& -i\beta&i\beta&\alpha&-\alpha&\beta&-\beta\\
   -c\alpha^2&-c\alpha^2&-c\beta^2&-c\beta^2&c\alpha^2& c\alpha^2& c\beta^2&c\beta^2\\
   ic\alpha^3&ic\alpha^3&-ic\beta^3&ic\beta^3&-c\alpha^3&c\alpha^3&-c\beta^3&  c\beta^3\\
   \gamma_1&\gamma_2&  \gamma_3& \gamma_4& \gamma_1&  \gamma_2&\gamma_3&\gamma_4\\
   1i\gamma_1\alpha& -1i\gamma_2\alpha&1i\gamma_3\beta&-1i\gamma_4\beta&-\gamma_5\alpha&\gamma_6\alpha&-\gamma_7\beta&\gamma_8\beta\\
   c\gamma_1\alpha^2&c\gamma_2\alpha^2& c\gamma_3\beta^2&c\gamma_4\beta^2&-c\gamma_5\alpha^2&-c\gamma_6\alpha^2&-c\gamma_7\beta^2&-c\gamma_8\beta^2\\
   -ic\gamma_1\alpha^3& ic\gamma_2\alpha^3& ic\gamma_3\beta^3& ic\gamma_4\beta^3 &-c\gamma_5\alpha^3&c\gamma_6\alpha^3& -c\gamma_7\beta^3&c\gamma_8\beta^3\\
\end{bsmallmatrix}
\end{equation}

\begin{equation}
\begin{aligned}
D = 
    \begin{bmatrix}
        e^{ik_1d}& 0& 0& 0\\
        0& e^{ik_2d}& 0& 0\\
        0& 0& e^{-ik_1d}&\\
        0& 0& 0& e^{-ik_2d}
    \end{bmatrix}
\end{aligned}    
\end{equation}

\begin{equation}
\begin{aligned}
    &\alpha= \left(a + 2*\overline{b}^2 + \sqrt{4*\overline{b}^4 + 2*\overline{b}^2*a}\right) ^{0.25}\\
    &\beta=\left(a + 2*\overline{b}^2 - \sqrt{4*\overline{b}^4 + 2*\overline{b}^2*a})\right)^{0.25}\\
    \end{aligned}
 \end{equation}   

\noindent
\begin{tabularx}{0.5\textwidth}{@{}XX@{}}
\begin{equation}
a=\sqrt{\dfrac{EI}{(pA)}}    
\end{equation} &
\begin{equation}
  C = \sqrt{\dfrac{E}{p}}   
\end{equation}\\

\begin{equation}
 \omega = \dfrac{2\pi wd}{C}   
\end{equation} &
\begin{equation}
b = \dfrac{2iw\omega}{C^2} 
\end{equation}\\

\begin{equation}
\overline{b}= \frac{b}{2i}
\end{equation}
\end{tabularx}
\end{flushleft}

In the equations above, $j$ is an indicator of each frequency, and $i$ is an indicator of the cross-sections.  $E, A, w$ denote the elastic (Young's) modulus, cross-sectional area, and width, respectively, and $\omega$ is the angular frequency. $C$ represents the wave, and $c$ is the effective elongation of the force. $T_{i,j}$ indicates the transfer matrix in specific cross-section for a specific frequency, while $\bar{T_{j}}$ indicates total transfer matrix of a particular frequency. 
\subsection{Surrogate Modeling}
Surrogate Modelling or metamodeling techniques are useful in optimization when the objective or constraint function(s) are computationally expensive. In this paper,in the 1D metamaterial design process, we consider constraints such as the frequency range between two successive peaks in transfer ratio,which ensures a fixed non-resonant frequency range, while the objective function -in this case mass of the inserts, is minimized. The Transfer Matrix Method is used to compute the frequency range . While computing the vibration characteristics using the TMM method is significantly more efficient than say a finite element analysis approach, its computational expense ($\sim$ 150 seconds on an Intel Core i7-9750H with 32GB system memory, to calculate axial, lateral and torsional vibration response) becomes substantial in the context of our needs -- where the model has to be used to generate multiple model-based sample optimizations. Hence, we leverage a surrogate modeling approach to substitute the transfer matrix method. Here, shallow feed-forward neural networks are used for this purpose.

\subsection{Optimization}
Optimizations are performed with two different objective functions to suit the needs of two different end goals from a vibration mitigation standpoint. The first objective function, specified in section \ref{sec: optimization}, is used to design an aperiodic metamaterial with very wide non-resonant frequency ranges. The results of this optimization are compared with specified baseline designs. 
Another optimization is done to generate samples for the invertible ANN, which is discussed in section \ref{sec:inv_des}. Since, invertible ANNs require a bijective function spaces (i.e., one-to-one mapping), constrained optimization is used to facilitate locally unique forward solutions.
The PSO algorithm by Chowdhury et al. \cite{chowdhury2013mixed} is used for optimization. Default PSO settings were used, which can be found in \cite{chowdhury2013mixed}.

\subsection{Inverse Mapping with Invertible Neural Networks}
Many engineering and materials design problems can be expressed as an inverse problem, where a well-posed forward mapping $f: x \rightarrow y$ exists, and we want to find the values of variables $x$, which corresponds to the desired value of a property $y$. The mapping $f: x \rightarrow y$ is well defined and easy to compute, but the inverse generally has to be found by solving an optimization problem. 
However, solving an optimization problem for every design query is inefficient and time-consuming, and does not lend well to on-demand design retrieval. Such \textbf{on-demand design retrieval} is useful in various contexts, such as i) conceptual design, ii) feasibility analysis, iii) fast retrieval of intermediate parameters or test of existence of feasible intermediate parameters in multi-scale or hierarchical design problems, and iv) creation of seeding databases, i.e., databases to effectively seed more complex metamaterial optimizations or design of experiments.

An invertible forward model ($f: x \rightarrow y$) will allow retrieval of the inverse ($f_{\texttt{inv}}: y \rightarrow x$), which is the role being played by invertible ANNs in this paper. Invertible networks or INNs \cite{inversenets} are a class of neural networks that have a unique architecture, which ensures their invertibility. While traditional neural networks, when applied to inverse problems, try to directly approximate the complex inverse problem \cite{inversenets}, INNs when trained on the well-understood forward problem, will capture the inverse model automatically. However, INNs can work only when the model is bijective. 
In this paper, we first train a surrogate model (from the TMM response) to decrease the computational cost of optimization, and later we use it to train our invertible neural network.  It is also possible to use the invertible neural network as a surrogate model, but as mentioned above, additional constraints are necessary to be able to use INNs in this manner, due to the non-unique nature of the inverse.
A handful of notable INN architectures have been reported in recent years \cite{mccann2017convolutional}, with preliminary applications to image reconstruction, parameter estimation, generative flow modeling. We implemented the Invertible Network architecture proposed in \cite{inversenets}; we chose this implementation as it offered an efficiently computable jacobian and ensure invetibility by preserving the non-singular nature of Jacobian. This implementation of INNs depends on the third set of variables called \textit{latent variables}. They represent the information lost in the forward mapping $x\rightarrow y$. The INN uses the relationship between these latent variables and the actual parameters in order to mitigate representation loss. Another key feature of this INN implementation is that the forward and the inverse models are trained jointly. 
Section \ref{sec:inv_des} and Table \ref{tab:set_optim} contains further details of our current implementation.

\section{Problem Definition}
\label{sec:prob_def}

\subsection{Design Using Aperiodic Inserts}
In this study, for passive vibration suppression, additional material is added to the drill string in the form of annular rings - the position and size of which are determined by optimization. Adding mass to the drill string affects the impedance response of the drill string, limiting the undesirable high transmission ratios to the stop band of the response. The number of inserts added is fixed at 10 for this study. This approach is similar to the one followed by Shaffer et al. \cite{alsaffar2018band}. The key difference is our usage of an aperiodic structure in contrast to the classical periodic structure used earlier. This leads to greater flexibility in the design of the drill string, while at the same time increasing the complexity of design space that must be searched (both in the forward and inverse problems). The increased flexibility not only offers increased performance, but also opportunities for reconfiguration during the lifetime of the system, due to the lack of dependence on a self-repeating (typical) metamaterial design. The designed pipe can be used for different vibration frequencies by merely adding or changing the position of the inserts. However, the additional inserts may introduce regions of higher stress concentration along the drill pipe. 

\begin{figure}[htbp]
  \includegraphics[width=\linewidth]{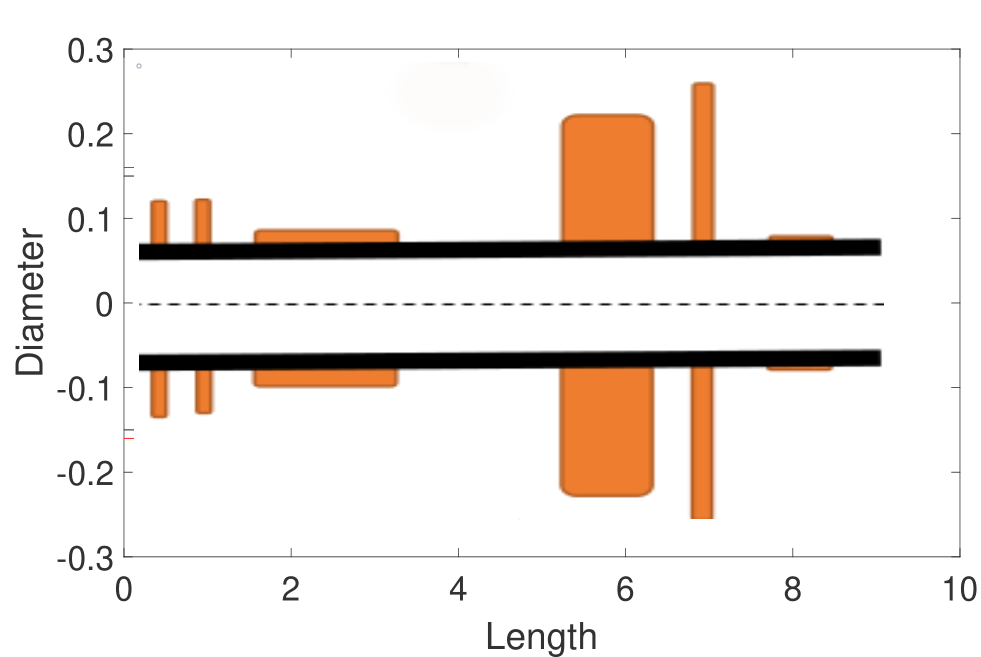}
  \caption{Structure of the drill pipe with inserts}
  \label{fig:flow}
\end{figure}

As a drill string is a very long structure with varying lengths (depending on the number of drill pipes connected), the design is performed on a more standardized structure -- a single drill pipe. Additionally, it is reasonable to expect the total vibration in the string will decrease if the vibration in every drill pipe decreases. The transmission ratios of the drill pipe are calculated by using TMM and constrained with a free-free boundary condition. 
Figure \ref{fig:flow} shows the design structure with the inserts (note that the radial dimension is enhanced to highlight the aperiodic nature of our designs). As seen from the figure, 
the aperiodic design is parameterized in terms of $D_i$, $W_{ring,i}$, $W_{No-ring,i}$, which are respectively the outer diameter of the $i$-th insert, the width of the $i$-th insert, and the width of the gap between the $i$-th and $i+1$-th inserts. The properties of the inserts and drill pipe are listed in table \ref{tab:set_prob}.

\begin{table}[!ht]
\caption{Drill Pipe properties}
\label{tab:set_prob}
\begin{center}
\begin{tabular}{|l|c|}
\hline
\large{\textbf{Pipe's Parameters}} & \large{\textbf{Value}}\\
\hline
\textbf{$L$}    &9 m \\
\textbf{$\rho$}    &1800 kg/m$^3$ \\
\textbf{$E$}    &193 GPa \\
\textbf{$G$}    &77.2 GPa \\
\textbf{$D_{in}$}    & 15 cm \\
\textbf{$D_{out}$}    &16 m \\

\hline

\large{\textbf{Insert's Parameters}} & \large{\textbf{Value}}\\
\hline
\textbf{$\rho$}    &1800 kg/m$^3$ \\
\textbf{$E$}    &193 GPa \\
\textbf{$G$}    &77.2 GPa \\
\textbf{$d$}    &  [ 16,32 ] cm \\
\textbf{$W_{ring}$}    &  [ 7.5,37.5 ] cm\\
\textbf{$W_{No-ring}$}    &  [ 0.15, 2.25 ] cm\\

\hline
\large{\textbf{Frequency Ranges}} & \large{\textbf{Value}}\\
\hline
\textbf{Axial}    &  [ 0.1, 800 ] Hz \\
\textbf{Torsional}    &  [ 0.1, 800 ] Hz\\
\textbf{Lateral}    &  [ 0.1, 10000 ] Hz\\
\hline
\end{tabular}
\end{center}
\end{table}


\subsection{Optimization formulation}
\label{sec: optimization}

 The optimization objective is to maximize the non-resonant frequency range and is formulated as:

\begin{equation}
\begin{split}
   &\underset{x}{\text{max:}} \omega_l-\omega_k, \ni \forall \omega \in (\omega_k, \omega_l) \sum_{\omega_k}^{\omega_l} \delta(x,\omega_i)= 0 \\
    &\delta(x,\omega_i) = 
    \begin{cases}
        1, & \text{if } \Psi(x,\omega_{i}) \geq \Psi(x,\omega_{i \pm 1}) \\
        0, & \text{otherwise}
    \end{cases} \\
    &\text{s.t: } \omega_l -\omega_k \geq \omega_c,\\
    &\omega_c = 
    \begin{cases}
       200 ~Hz, & \text{for Axial and Torsional Vibrations}  \\
        1000 ~Hz, & \text{For Lateral Vibrations}
    \end{cases} 
\end{split}
\end{equation}
\begin{flushleft}
\begin{equation}
    \begin{split}
	&D_{\min}  \leq D_i \leq D_{\max} , \ D_i \in \mathbb{R}\\
	&W_{ring,\min}  \leq W_{ring,i} \leq W_{ring,\max} , \ W_{ring,i} \in \mathbb{R}\\
	&W_{No-ring,\min}  \leq W_{No-ring} \leq W_{No-ring,\max} , \ W_{No-ring,i} \in \mathbb{R}\\	
	\end{split}
\end{equation}
\end{flushleft}
%
The vector of design variables,
\[x=\{D_i,W_{ring,i},W_{No-ring,i}, \forall i=1,2,\ldots,n\}\] contains the outer diameter of the insert, thickness of the insert, and the distance between successive inserts. For our case studies, we consider a fixed number of $n=10$ inserts.







Initial numerical (optimization) experiments showed that the problem is highly multi-modal, as a result of which a global optimizer such as PSO was deemed suitable to perform the optimizations.
\subsubsection*{Non-Resonant Frequency Range}
In this study, the axial and torsional vibrations were analyzed over a frequency range of 0.1-800 Hz, while the lateral vibrations were analyzed over a frequency range of 0.1-10000 Hz.
In order to find the transfer ratio across the domain ,the entire frequency range is discretized into 80 frequency points and the transmission ratios are calculated at those frequency points. This discretization is done in order to reduce the computational cost. The non-resonant frequency range is found by analyzing the transmission ratio values over the 80 points and identifying the resonant peaks. The largest range between two consecutive resonant highs was considered for optimization purposes.

\subsubsection*{Surrogate Modelling}
For each of the 80 discretized frequencies, a surrogate model is trained that can represent the attenuation of the 6 DoF's of the system. Therefore a total of 80 multi-input-6-output ANN models (with 100 hidden neurons in a single layer) were trained, which showed more promising results compared to training a multi-input-6$\times$80-output model. In order to train these models, 24,000 samples (of 30 design variables $D_{1, \ldots, 10}, W_{ring, 1, \ldots, 10}, W_{No-ring, 1, \ldots, 10} $ were generated. 
Table \ref{tab:set_optim} lists the model structure and settings used here. 

\section{Inverse Design}
\label{sec:inv_des}
For a structural dynamics problem such as the one dealt with in this paper, the forward problem consists of finding the frequency response for a given structure, which can be represented as:
\begin{equation}
     \dv[2]{}{X}\left[ EI\dv[2]{u}{X}\right] = -M \dv[2]{u}{t} + p(X)
     \label{eq:DYN_EQ}
\end{equation}
Where E is the elastic modulus, I is the moment of inertia of the structure,p(x) is the external load on the system and M is the mass. 
The frequency response is obtained by solving the above PDE, in this case, using the TMM method. In the context of our vibration mitigation goals, we are interested in determining the range of frequencies between which no resonance occurs, and thus the ``PDE solution of interest" is represented in terms of a counter ($h(x,\omega_1,\omega_2)$) that counts the number of peaks in the frequency response between $\omega_1$ and $\omega_2$.  The inverse problem can thus be formulated as the following PDE-constrained optimization problem, as shown below. 
\begin{equation}
    \begin{aligned}
   &\underset{x}{\min} \hspace{10pt} m(x) = \rho \times \sum_{i=1}^{10}\frac{\pi}{4}(x_{d_i}^2-D_i^2)x_{w_i}\\
     &\text{s.t.\hspace{10pt}}
 h(x,\omega_1,\omega_2) =
   \sum_{\omega_1}^{\omega_2} \delta(x,\omega_i)= 0 
     , {\rm{where}}
    \end{aligned}
\end{equation}
\begin{equation}
    \delta(x,\omega_i) = 
    \begin{cases}
        1, & \text{if }   \Psi(x,\omega_{i}) \geq \Psi(x,\omega_{i \pm 1})  \\
        0, & \text{otherwise}
    \end{cases}
\end{equation} 
{\rm{where}}, $\Psi(x,\omega_i) = \dfrac{u}{p(X)}$. 
In the above equations, $\omega_1$ and $\omega_2$ are the frequencies that bound a frequency range where no resonance occurs (which we call the ``non-resonant frequency range"), and delta is a binary operator signifying if there is a peak in transfer ratio at a given frequency.

As mentioned earlier, the INN needs a bijective function mapping in order to ensure the non-singular nature of the Jacobian which then ensures invertibility. However, the current problem is not bijective by itself (as is common with inverse problems) since there can be more than one metamaterial (drill pipe) configurations that provides similar non-resonant frequency range.
Therefore, to impose uniqueness, samples are generated by minimizing mass for varying windows of non-resonance, as shown in the above optimization formulation. The premise here being that any local minimum in terms of mass ensures a unique solution in its neighborhood. 
\section{Results and Discussion}
\label{sec:res}

\textcolor{black}{Table \ref{tab:set_optim} lists the settings of the optimization problem. }

\begin{table}[!ht]
\caption{Optimization and Learning Settings}
\label{tab:set_optim}
\begin{center}
\begin{tabular}{|l|c|}

\hline
{\textbf{Surrogate ANN Learning Setting}} & {\textbf{Value}}\\
\hline
\textbf{Input Size}    & 30 \\
\textbf{Output Size}    &  6\\
\textbf{Hidden Layers}    &  1\\
\textbf{Activation Function}    & tanh \\
\textbf{Maximum Iterations}    & 200 \\
\textbf{Optimizer}    & SGDM \\

\hline
{\textbf{PSO Optimization settings}} & {\textbf{Value}}\\
\hline
\textbf{Design Variables Size}    & 30 \\
\textbf{Population}    & 300 \\
\textbf{Maximum Iterations}    &  50\\

\hline
{\textbf{Inverse ANN Learning Setting}} & {\textbf{Value}}\\
\hline
\textbf{Input Size}    & 30 \\
\textbf{Output Size}    & 2(+28 \textit{latent variables}) \\
\textbf{Hidden Layers}    &  2\\
\textbf{Activation Function}    & leaky ReLu \\
\textbf{Maximum Iterations}    &  1500\\
\textbf{Learning Rate}    &  $10^{-3}$ to $0.02\times 10^{-3}$\\
\textbf{Optimizer}    & ADAM \\

\hline

\end{tabular}
\end{center}
\end{table}

\subsection{Optimization Results}

\begin{figure}[!htbp]
\begin{center}
      \includegraphics[width=7cm,height=6cm]{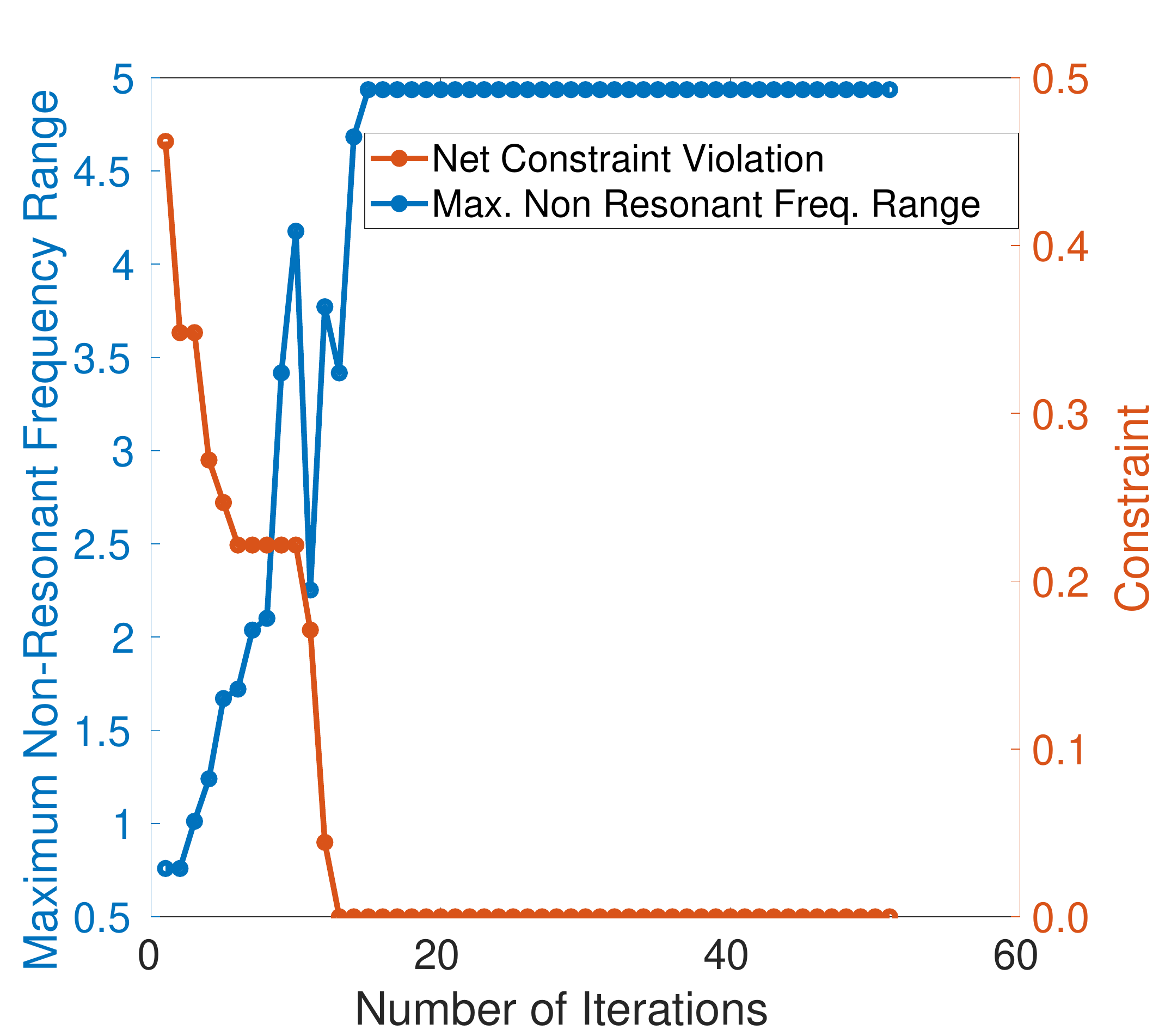}
  \caption{Maximization of Non-Resonant Frequency Range optimization convergence history}
  \label{fig:MASS_obj}
  \end{center}
\end{figure}

The pursuit of the optimal design was conducted using a population of 300 particles and a maximum of 50 iterations in the PSO algorithm. Figure \ref{fig:MASS_obj}  shows the convergence history of the first optimization.  
The convergence history shows that the system successfully generated the required non-resonant frequency range in 3 DoFs, but the total mass remained large, which was expected for this design. 

\begin{figure}[!htbp]
\begin{center}
      \includegraphics[width=7.5cm,height=6cm]{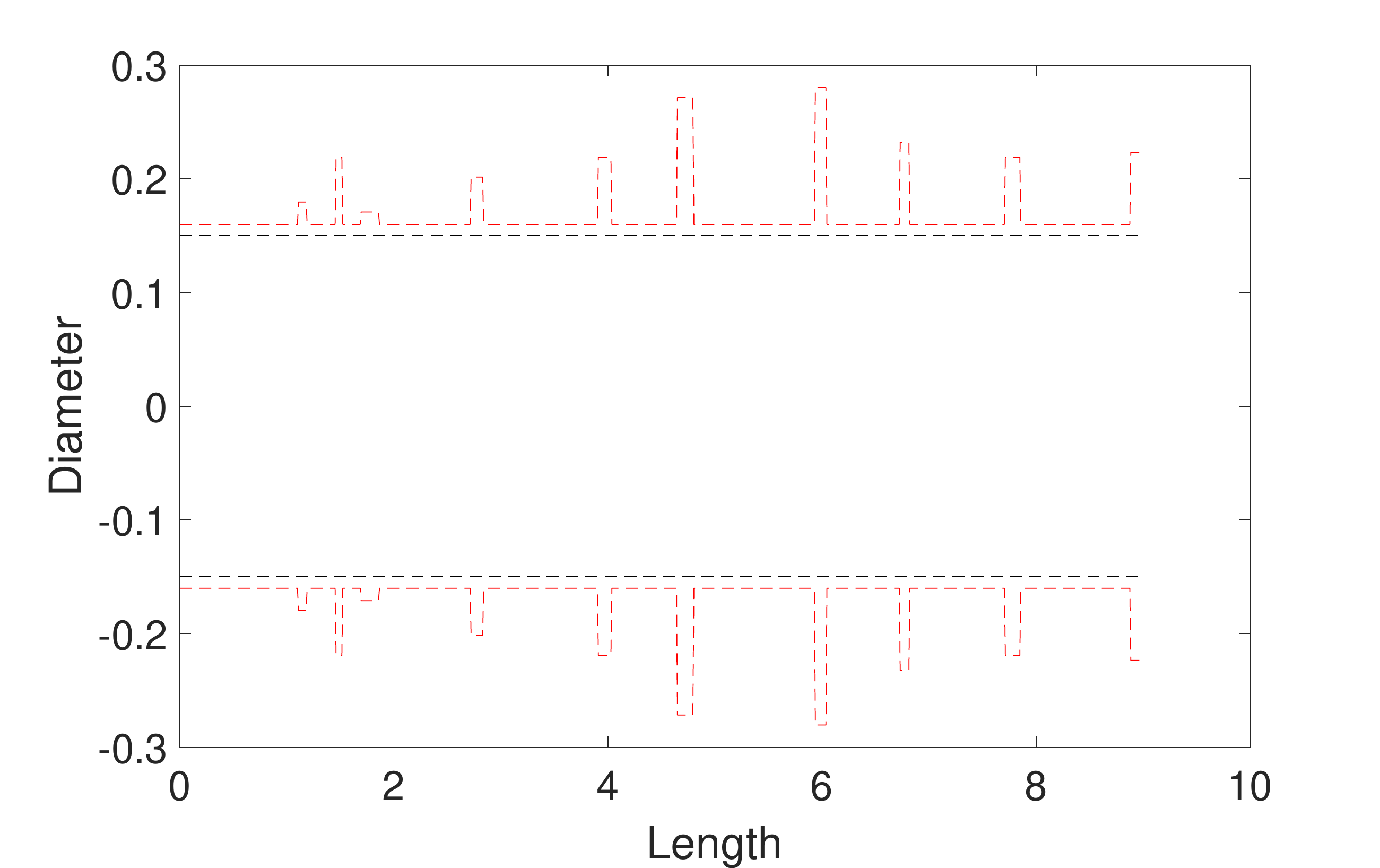}
  \caption{ Baseline Design Geometry}
  \label{fig:shape_base}
  \end{center}
\end{figure}

\begin{figure}[!htbp]
    \begin{center}
      \includegraphics[width=7.5cm,height=6cm]{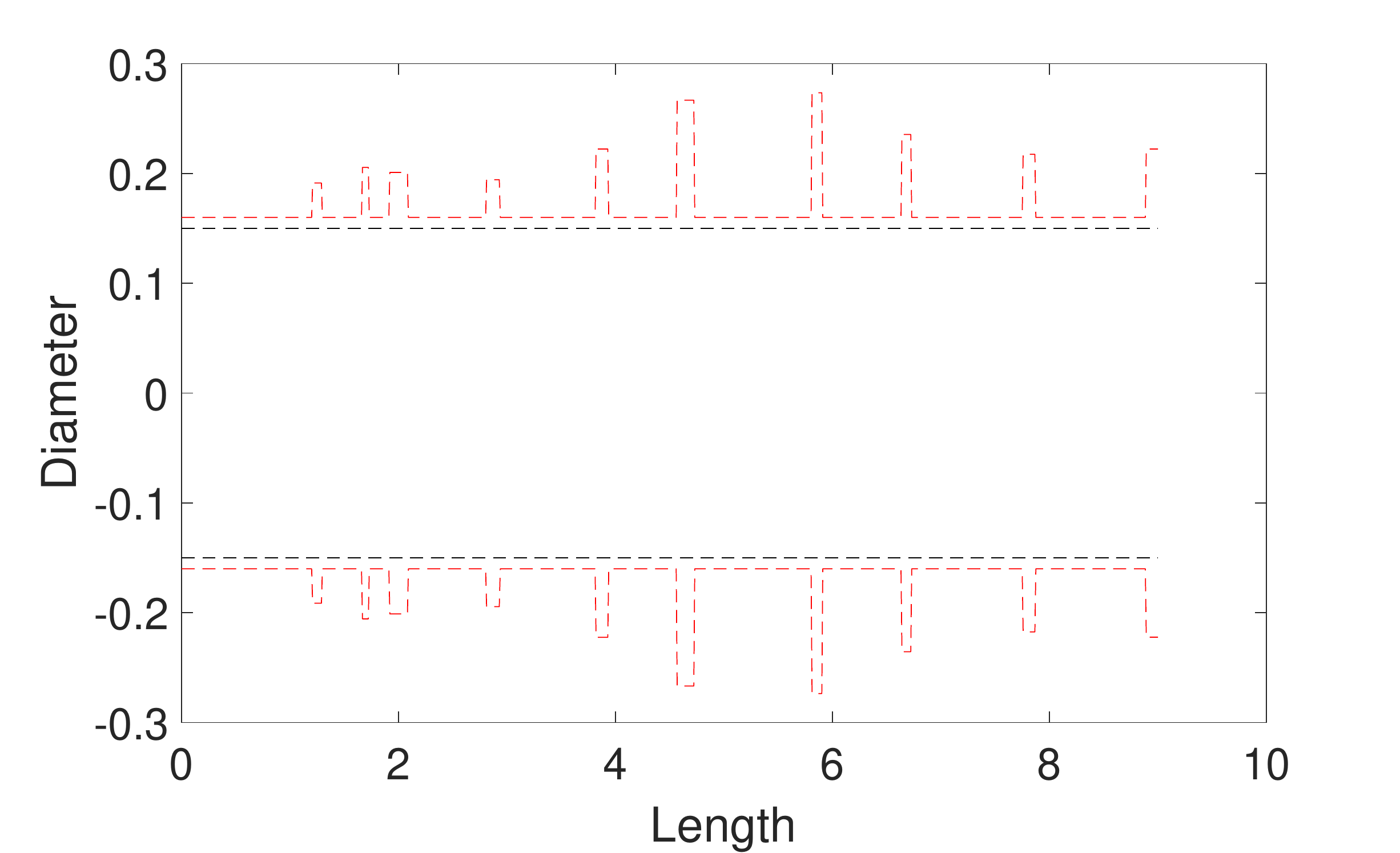}
  \caption{ Optimal Design Geometry}
  \label{fig:shape_mass}
  \end{center}
\end{figure}

\begin{figure}
\centering
\begin{subfigure}{.5\textwidth}
  \centering
  \includegraphics[width=7cm,height=6cm]{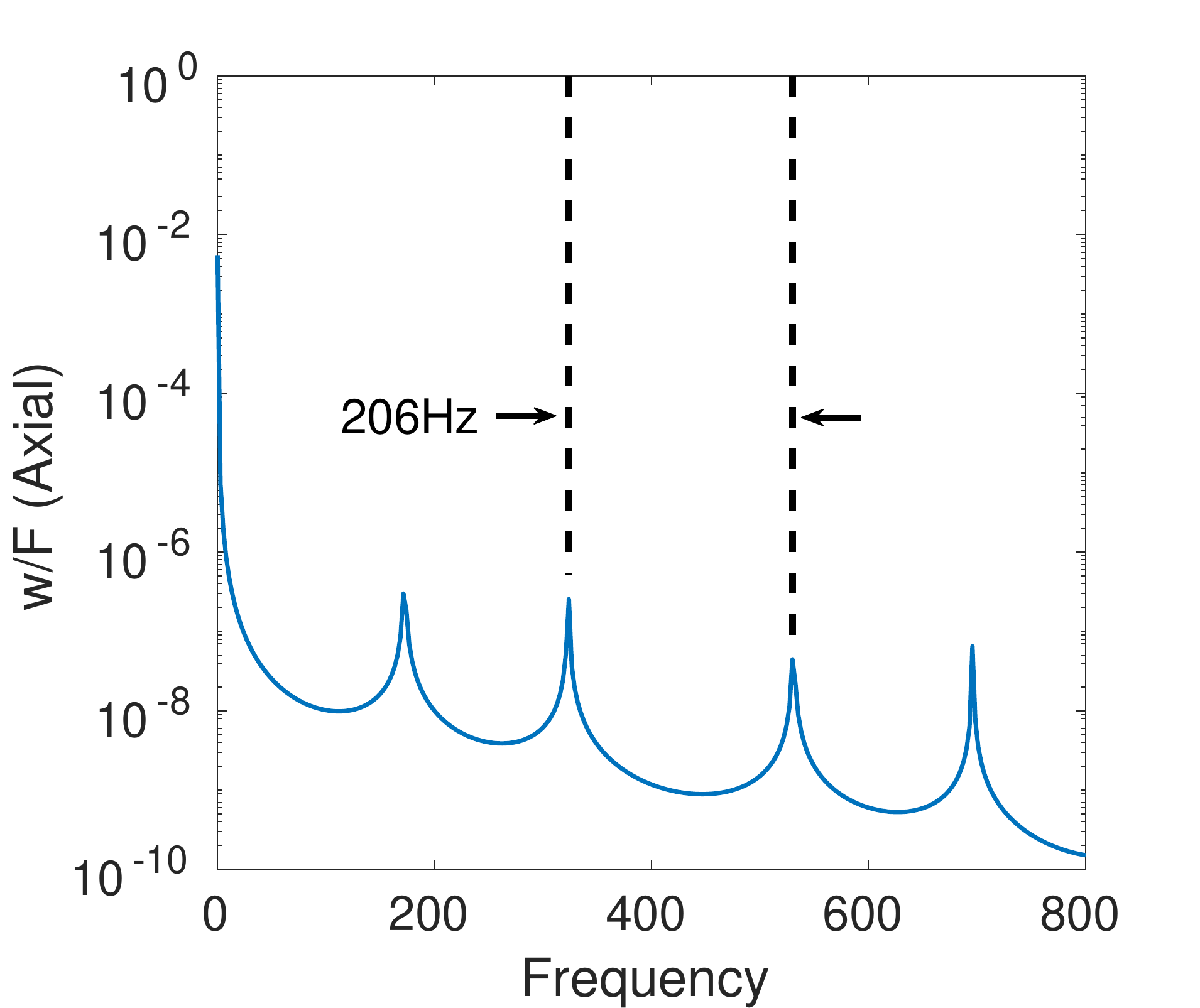}
  \caption{Axial Vibration}
  \label{fig:tf_base_1}
\end{subfigure}%
\\
\begin{subfigure}{.5\textwidth}
  \centering
  \includegraphics[width=7cm,height=6cm]{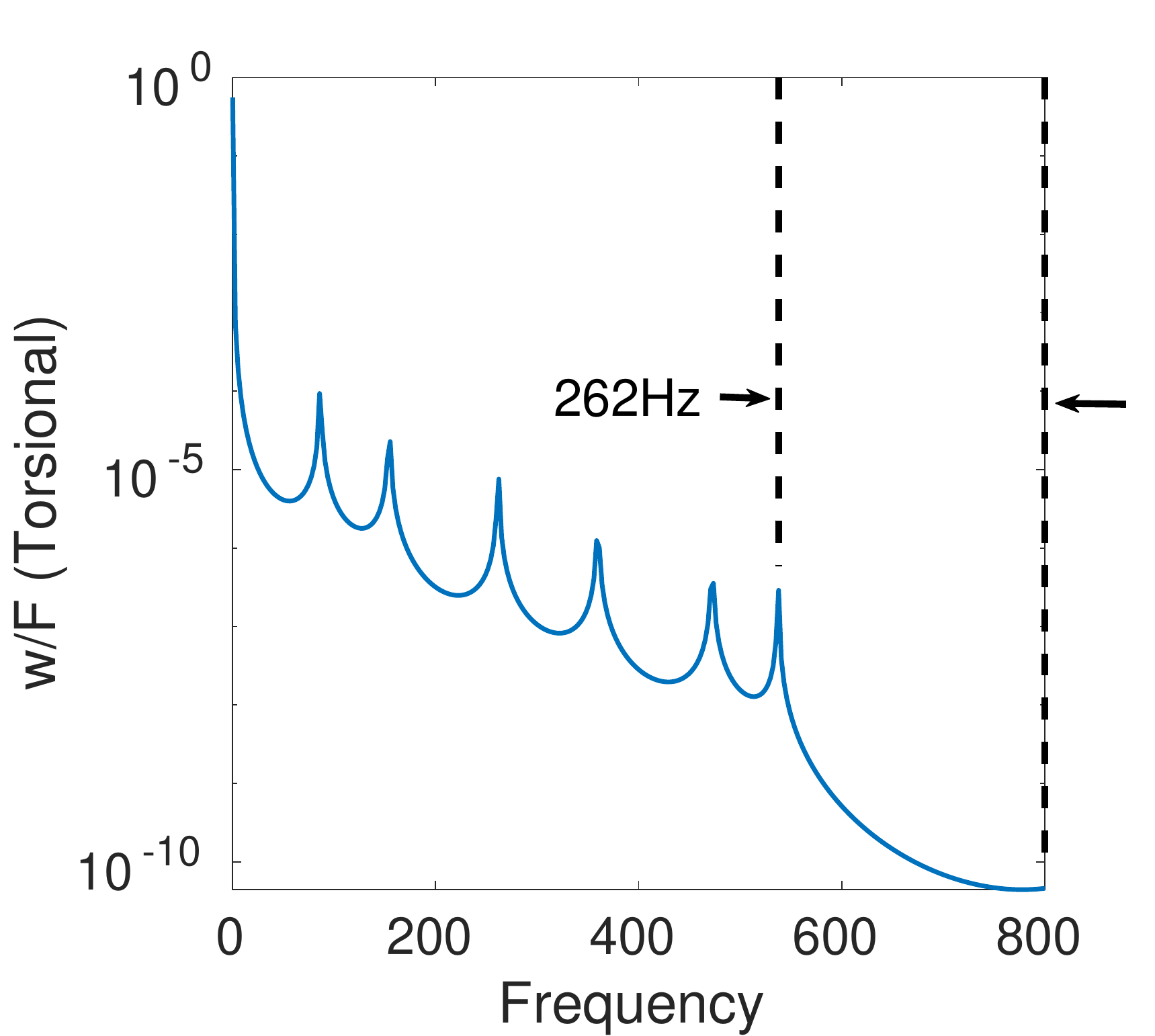}
  \caption{Torsional Vibration}
  \label{fig:tf_base_2}
\end{subfigure}
\\
\begin{subfigure}{.5\textwidth}
  \centering
  \includegraphics[width=7cm,height=6cm]{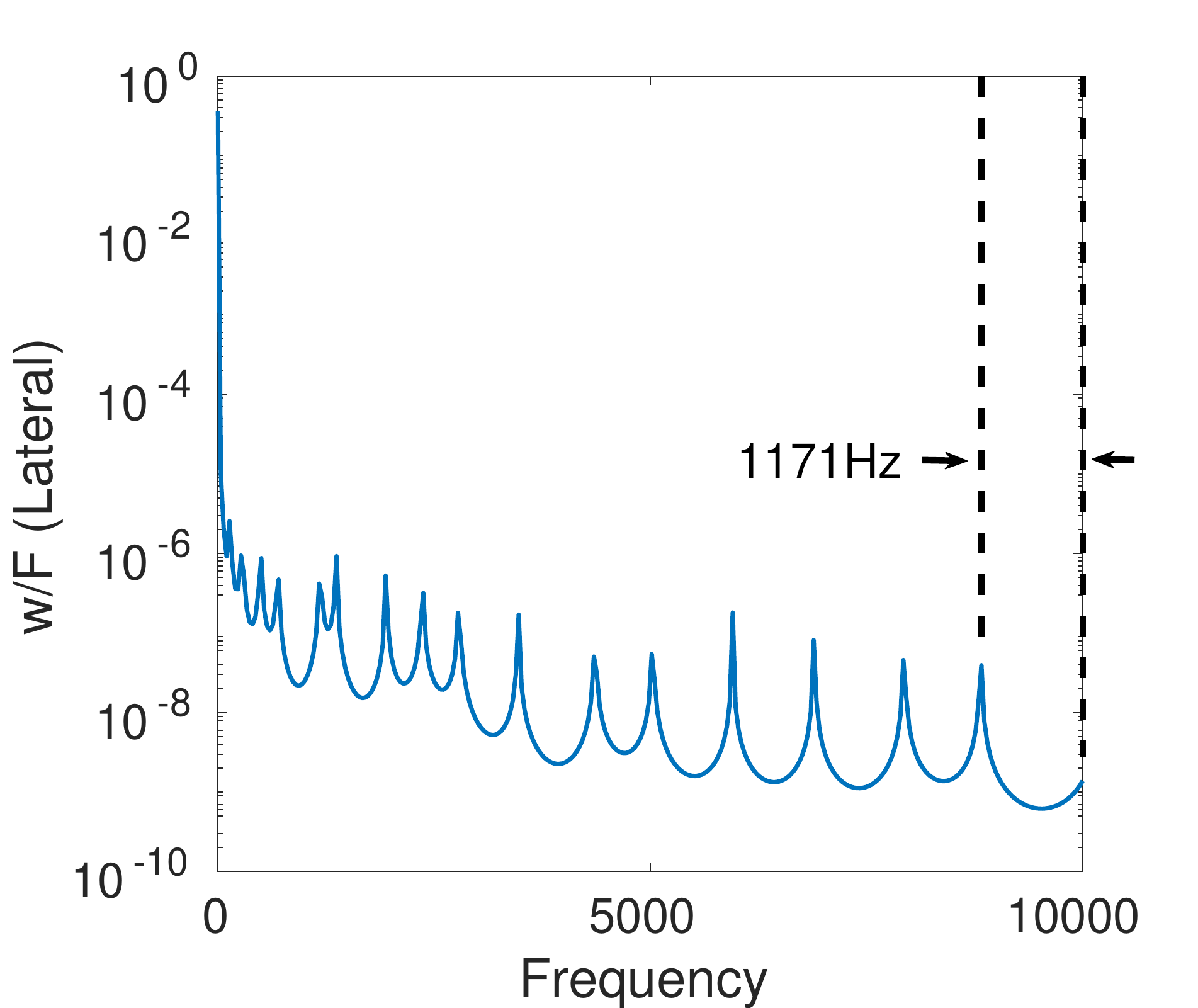}
  \caption{Lateral Vibration}
  \label{fig:tf_base_3}
\end{subfigure}
\caption{Baseline Design Frequency Response}
\label{fig:tf_base}
\end{figure}

\begin{figure}
\centering
\begin{subfigure}{.5\textwidth}
  \centering
  \includegraphics[width=7cm,height=6cm]{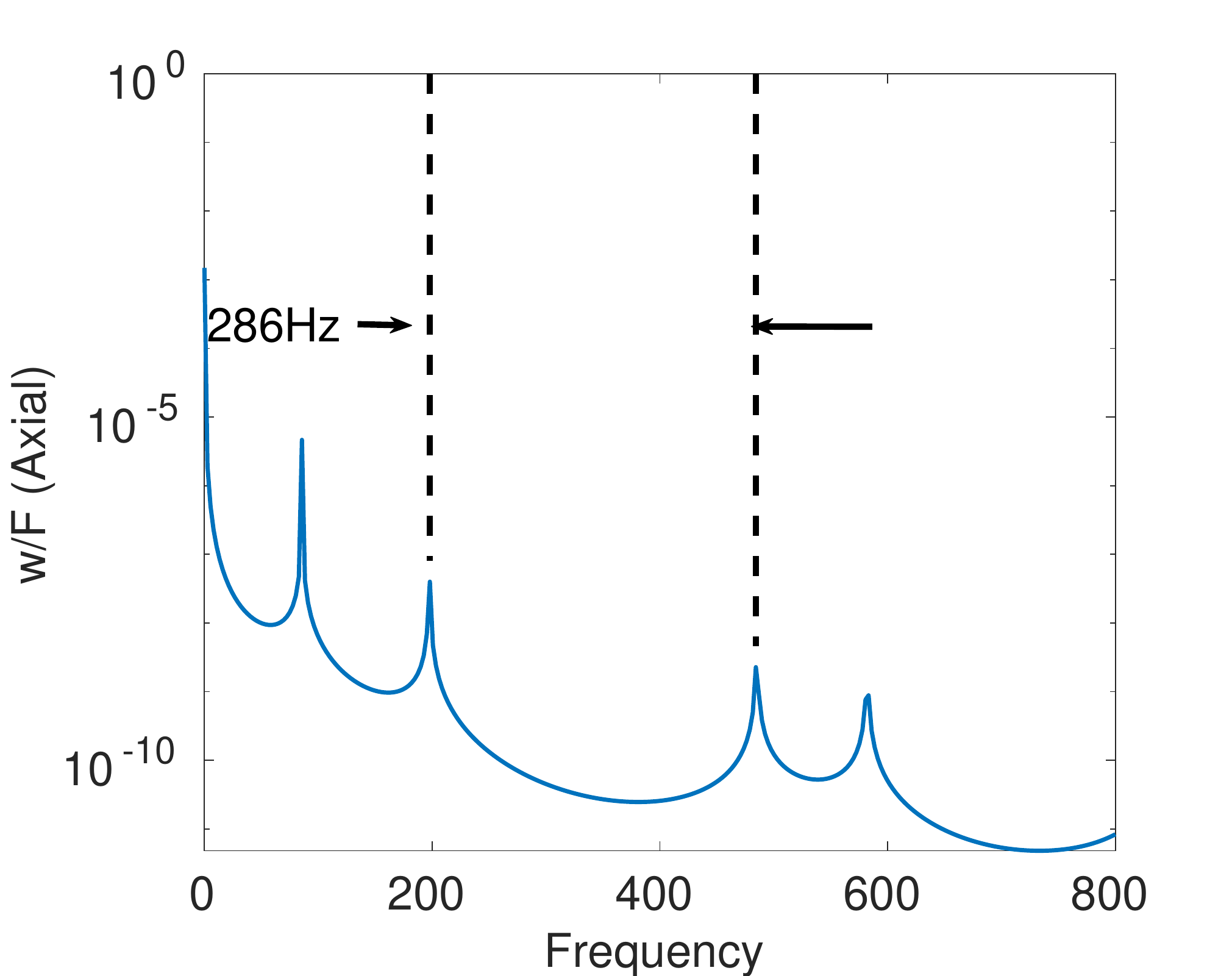}
  \caption{Axial Vibration}
  \label{fig:tf_mass_1}
\end{subfigure}%
\\
\begin{subfigure}{.5\textwidth}
  \centering
  \includegraphics[width=7cm,height=6cm]{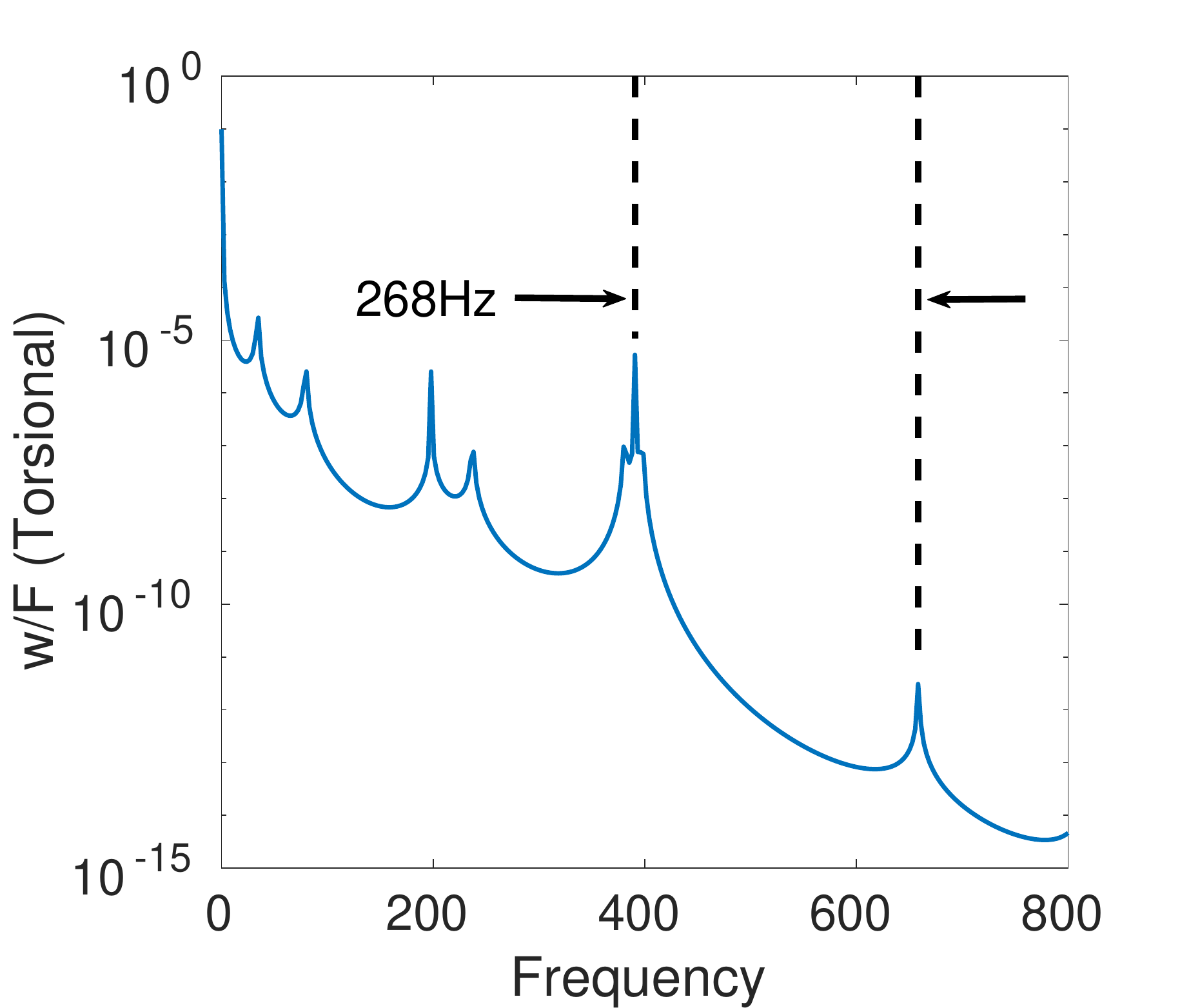}
  \caption{Torsional Vibration}
  \label{fig:tf_mass_2}
\end{subfigure}
\\
\begin{subfigure}{.5\textwidth}
  \centering
  \includegraphics[width=7cm,height=6cm]{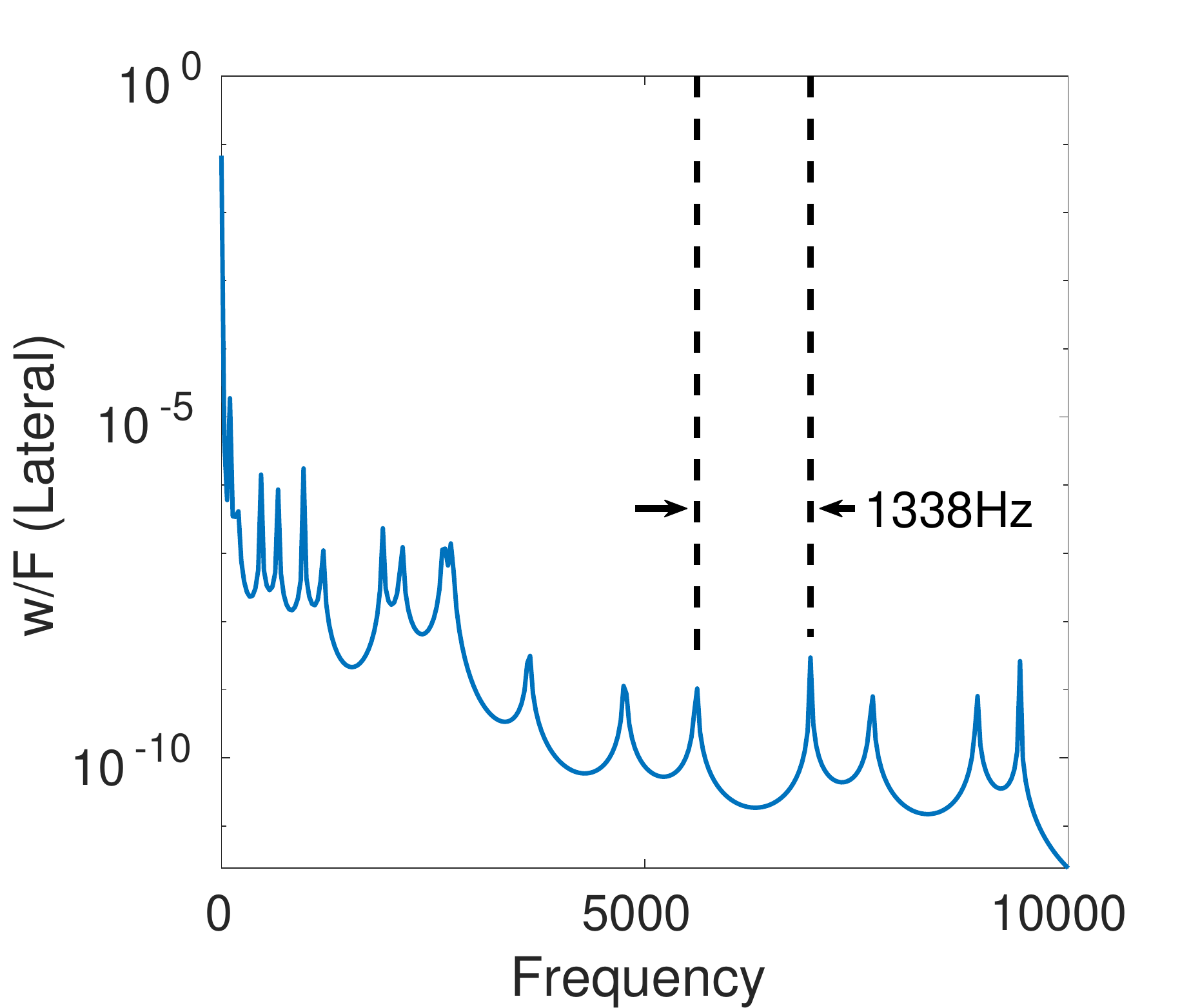}
  \caption{Lateral Vibration}
  \label{fig:tf_mass_3}
\end{subfigure}
\caption{Optimal Design Frequency Response}
\label{fig:tf_mass}
\end{figure}

Figure \ref{fig:shape_base} shows a initial randomly generated structure from optimization process, a representative baseline design. Figure  \ref{fig:shape_mass} shows the actual design of the drill pipe and Figs. \ref{fig:tf_base_1} to \ref{fig:tf_base_3} and \ref{fig:tf_mass_1} to \ref{fig:tf_mass_3} show the transfer ratios for different vibration modes for the baseline design and optimal design, respectively. The substantial improvement in the axial vibration non-resonant frequency range is readily evident, majorly attributed to design changes to the leftmost 3 inserts.   

\subsection{Inverse Design}
\begin{figure}
\centering
\begin{subfigure}{.5\textwidth}
  \includegraphics[width=8cm,height=6cm]{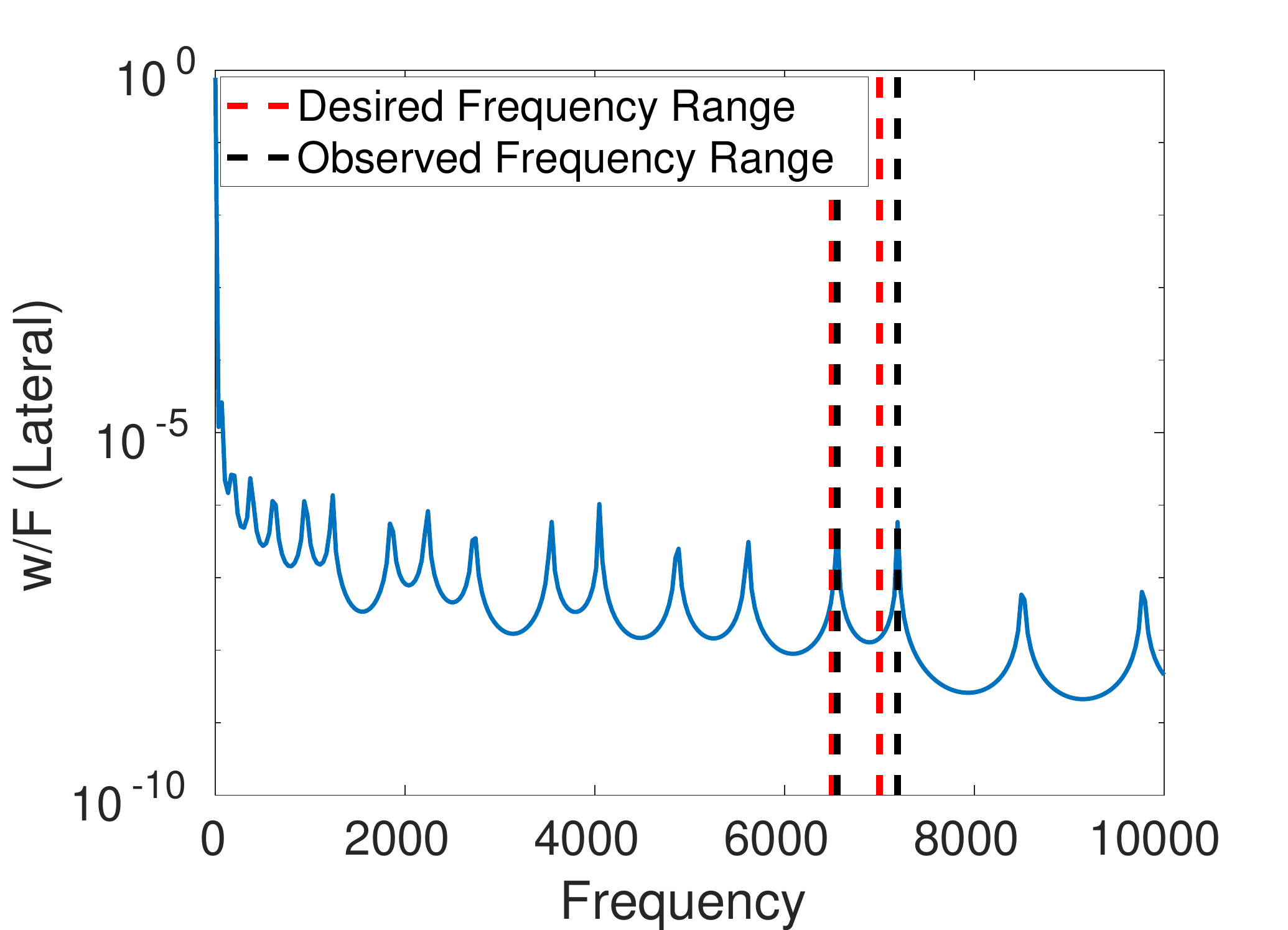}
  \caption{Frequency Response of Inverse Design Generated by INN}
  \label{fig:inndesign}
\end{subfigure}\\
\begin{subfigure}{.5\textwidth}
  \includegraphics[width=8cm,height=6cm]{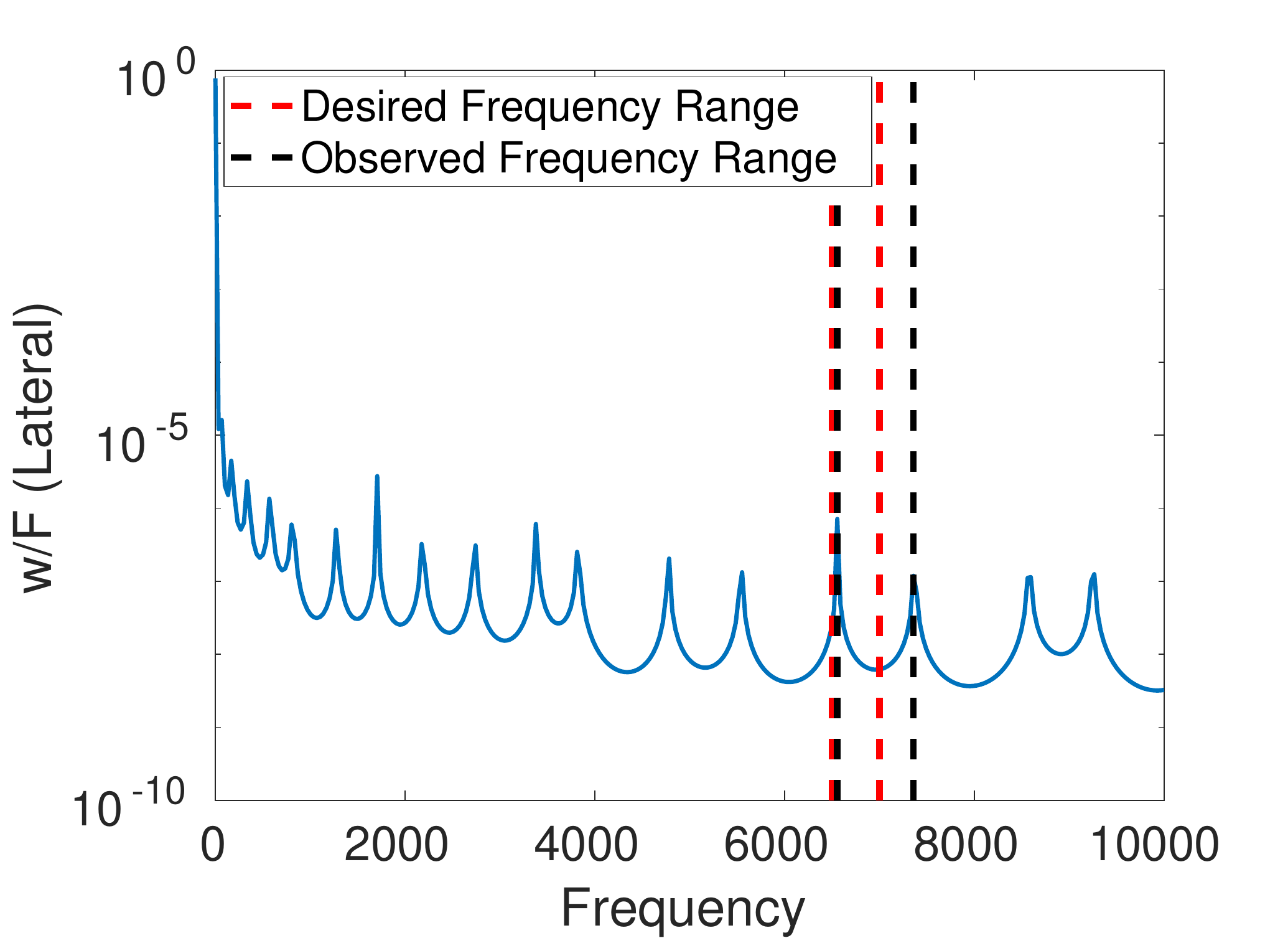}
  \caption{Frequency Response of Inverse Design Generated by Optimization}
  \label{fig:optdesign}
\end{subfigure}\\
\begin{subfigure}{.5\textwidth}
  \includegraphics[width=8cm,height=6cm]{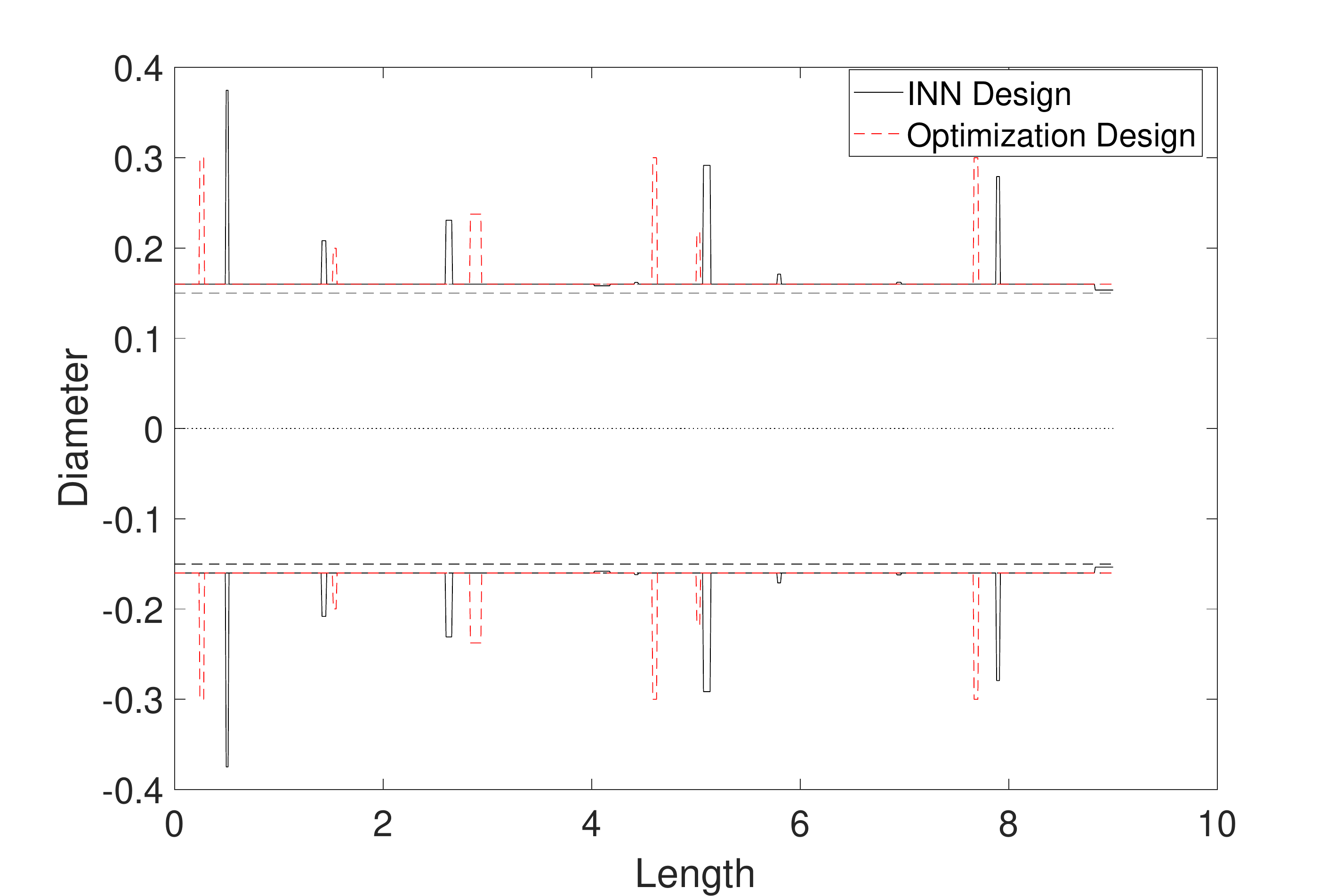}
  \caption{Shape of Drill Pipe Obtained from INN}
  \label{fig:inv_shape1}
\end{subfigure}\\ 
\caption{Comparison of Inverse Design Results}
\end{figure}

As mentioned earlier, the INN was trained on samples generated from optimization. For this purpose, 304 samples were generated with four different frequency band sizes over the total frequency range. The inputs to the INN are the design variables and outputs are the limits of the non-resonant frequency range. The trained model can be used to recover the design of the drill pipe with inserts, for a given upper and lower bound of the non-resonant frequency range. The training mean squared error (MSE) of the INN was found to be 0.09. Testing error over unseen samples was found to be 0.13 in terms of MSE. Here we show the performance of the trained INN on a test sample. Specifically, Fig. \ref{fig:inndesign} shows the design generated by the INN when the limits to the non-resonant frequency range are 6500 and 7000 Hz. Figure \ref{fig:optdesign} shows the result obtained from optimization for the same non-resonant range. \\

These results indicate that the Inverse Network outputs a design that has a frequency range wider than the desired range, while the forward optimization produces a range that is almost equal to the required bounds.
 While both solutions satisfy the non-resonant frequency range constraint, the design obtained from optimization weighs 249 kg while the design generated by the weighs INN 257 kg. 
This discrepancy is not surprising and could be attributed to either the error in the INN model, or more likely to the multi-modal nature of the forward optimization problem. More comprehensive statistical analysis, augmented by error propagation approaches will allows shedding more light on this discrepancy and its implication for on-demand design.

\section{Conclusion}
\label{sec:cons}
In this paper, an inverse design and surrogate-based optimization approach for designing aperiodic metamaterials are studied. In order to explore the metamodel based forward design and the novel inverse design approach (based on on invertible neural nets or INNs), a passive vibration suppression problem for drilling pipes is investigated.
 First, the Transfer Matrix Method is used to generate a transmission ratio for Axial, Torsional, and Lateral directions of 1-D unit cells. In our problem, we insert cylindrical rings on a drilling pipe to decrease its vibration transmission ratio and generate non-resonance frequency bands. Then multiple ANNs are used as forward surrogate models to substitute the Transfer Matrix Method. These models were used for surrogate-based optimization and helped in making sample generation tractable, where these samples are to be used for training the invertible neural network. Once trained, the INNs can provide near instantaneous retrieval of designs, given desired non-resonant freqneucy response properties, which has tremendous value for conceptual design, database generation and other applications. Testing performance and validation showed the effectiveness of the inverse design learning approach. Since, this was our first foray, i.e., a proof of concept study in exploring the benefits of invertible neural nets in inverse design of metamaterials, we made several simplifying assumptions. This included considering a 1D system (albeit with multiple coupled, this complex, vibration modes), fixed number of cells, and using optimizations to generate samples apriori to train the invertible neural nets. Many of these assumptions can be handled in the future, e.g., by incorporating smarter sampling techniques, which will help further establish the true potential of such inverse design methods, in the development of complex metamaterial systems. 
\bibliographystyle{IEEEtran}
\bibliography{bibliography}

\begin{thebibliography}{10}
\providecommand{\url}[1]{#1}
\csname url@samestyle\endcsname
\providecommand{\newblock}{\relax}
\providecommand{\bibinfo}[2]{#2}
\providecommand{\BIBentrySTDinterwordspacing}{\spaceskip=0pt\relax}
\providecommand{\BIBentryALTinterwordstretchfactor}{4}
\providecommand{\BIBentryALTinterwordspacing}{\spaceskip=\fontdimen2\font plus
\BIBentryALTinterwordstretchfactor\fontdimen3\font minus
  \fontdimen4\font\relax}
\providecommand{\BIBforeignlanguage}[2]{{%
\expandafter\ifx\csname l@#1\endcsname\relax
\typeout{** WARNING: IEEEtran.bst: No hyphenation pattern has been}%
\typeout{** loaded for the language `#1'. Using the pattern for}%
\typeout{** the default language instead.}%
\else
\language=\csname l@#1\endcsname
\fi
#2}}
\providecommand{\BIBdecl}{\relax}
\BIBdecl

\bibitem{fok2008acoustic}
L.~Fok, M.~Ambati, and X.~Zhang, ``Acoustic metamaterials,'' \emph{MRS
  bulletin}, vol.~33, no.~10, pp. 931--934, 2008.

\bibitem{albabaa2017formation}
H.~Al~Ba'ba'a, M.~Nouh, and T.~Singh, ``Formation of local resonance band gaps
  in finite acoustic metamaterials: A closed-form transfer function model,''
  \emph{Journal of Sound and Vibration}, vol. 410, pp. 429--446, 2017.

\bibitem{chronopoulos2017enhanced}
D.~Chronopoulos, I.~Antoniadis, and T.~Ampatzidis, ``Enhanced acoustic
  insulation properties of composite metamaterials having embedded negative
  stiffness inclusions,'' \emph{Extreme Mechanics Letters}, vol.~12, pp.
  48--54, 2017.

\bibitem{maruyama2010analysis}
T.~Maruyama, T.~Furuno, Y.~Oda, J.~Shen, and T.~Ohya, ``Analysis and design of
  metamaterial reflectarray using combination of multilayer
  mushroom-structure,'' in \emph{2010 IEEE Antennas and Propagation Society
  International Symposium}.\hskip 1em plus 0.5em minus 0.4em\relax IEEE, 2010,
  pp. 1--4.

\bibitem{buckmann2012tailored}
T.~B{\"u}ckmann, N.~Stenger, M.~Kadic, J.~Kaschke, A.~Fr{\"o}lich,
  T.~Kennerknecht, C.~Eberl, M.~Thiel, and M.~Wegener, ``Tailored 3d mechanical
  metamaterials made by dip-in direct-laser-writing optical lithography,''
  \emph{Advanced Materials}, vol.~24, no.~20, pp. 2710--2714, 2012.

\bibitem{yoon2016challenges}
G.~Yoon, I.~Kim, and J.~Rho, ``Challenges in fabrication towards realization of
  practical metamaterials,'' \emph{Microelectronic Engineering}, vol. 163, pp.
  7--20, 2016.

\bibitem{sui2015lightweight}
N.~Sui, X.~Yan, T.-Y. Huang, J.~Xu, F.-G. Yuan, and Y.~Jing, ``A lightweight
  yet sound-proof honeycomb acoustic metamaterial,'' \emph{Applied Physics
  Letters}, vol. 106, no.~17, p. 171905, 2015.

\bibitem{mamaghani2016vibration}
A.~E. Mamaghani, S.~Khadem, and S.~Bab, ``Vibration control of a pipe conveying
  fluid under external periodic excitation using a nonlinear energy sink,''
  \emph{Nonlinear Dynamics}, vol.~86, no.~3, pp. 1761--1795, 2016.

\bibitem{liu2008three}
N.~Liu, H.~Guo, L.~Fu, S.~Kaiser, H.~Schweizer, and H.~Giessen,
  ``Three-dimensional photonic metamaterials at optical frequencies,''
  \emph{Nature materials}, vol.~7, no.~1, pp. 31--37, 2008.

\bibitem{gansel2009gold}
J.~K. Gansel, M.~Thiel, M.~S. Rill, M.~Decker, K.~Bade, V.~Saile, G.~von
  Freymann, S.~Linden, and M.~Wegener, ``Gold helix photonic metamaterial as
  broadband circular polarizer,'' \emph{Science}, vol. 325, no. 5947, pp.
  1513--1515, 2009.

\bibitem{chen2012metamaterials}
T.~Chen, S.~Li, and H.~Sun, ``Metamaterials application in sensing,''
  \emph{Sensors}, vol.~12, no.~3, pp. 2742--2765, 2012.

\bibitem{albabaa2019emergence}
H.~A. Ba'ba'a and M.~Nouh, ``Emergence of pseudo-phononic gaps in periodically
  architected pendulums,'' \emph{arXiv preprint arXiv:1904.01396}, 2019.

\bibitem{nouh2019control}
M.~Nouh \emph{et~al.}, ``Control of spatial wave profiles in finite lattices of
  repelling magnets,'' \emph{Journal of Dynamic Systems, Measurement, and
  Control}, vol. 141, no.~11, 2019.

\bibitem{ge2007ga}
Y.~Ge and K.~Esselle, ``Ga/fdtd technique for the design and optimisation of
  periodic metamaterials,'' \emph{IET microwaves, antennas \& propagation},
  vol.~1, no.~1, pp. 158--164, 2007.

\bibitem{nouh2014vibration}
M.~Nouh, O.~Aldraihem, and A.~Baz, ``Vibration characteristics of metamaterial
  beams with periodic local resonances,'' \emph{Journal of Vibration and
  Acoustics}, vol. 136, no.~6, 2014.

\bibitem{nouh2015wave}
------, ``Wave propagation in metamaterial plates with periodic local
  resonances,'' \emph{Journal of Sound and Vibration}, vol. 341, pp. 53--73,
  2015.

\bibitem{hussein2006multiobjective}
M.~I. Hussein, K.~Hamza, G.~M. Hulbert, R.~A. Scott, and K.~Saitou,
  ``Multiobjective evolutionary optimization of periodic layered materials for
  desired wave dispersion characteristics,'' \emph{Structural and
  Multidisciplinary Optimization}, vol.~31, no.~1, pp. 60--75, 2006.

\bibitem{yang2017aperiodic}
Q.~Yang, X.~Chen, Y.~Li, X.~Zhang, Y.~Xu, Z.~Tian, C.~Ouyang, J.~Gu, J.~Han,
  and W.~Zhang, ``Aperiodic-metamaterial-based absorber,'' \emph{APL
  Materials}, vol.~5, no.~9, p. 096107, 2017.

\bibitem{bostanabad2019globally}
R.~Bostanabad, Y.-C. Chan, L.~Wang, P.~Zhu, and W.~Chen, ``Globally approximate
  gaussian processes for big data with application to data-driven metamaterials
  design,'' \emph{Journal of Mechanical Design}, vol. 141, no.~11, 2019.

\bibitem{matthews2016hierarchical}
J.~Matthews, T.~Klatt, C.~Morris, C.~C. Seepersad, M.~Haberman, and D.~Shahan,
  ``Hierarchical design of negative stiffness metamaterials using a bayesian
  network classifier,'' \emph{Journal of Mechanical Design}, vol. 138, no.~4,
  2016.

\bibitem{morris2018design}
C.~Morris, L.~Bekker, M.~R. Haberman, and C.~C. Seepersad, ``Design exploration
  of reliably manufacturable materials and structures with applications to
  negative stiffness metamaterials and microstereolithography,'' \emph{Journal
  of Mechanical Design}, vol. 140, no.~11, 2018.

\bibitem{han2012surrogate}
Z.-H. Han, K.-S. Zhang \emph{et~al.}, ``Surrogate-based optimization,''
  \emph{Real-world applications of genetic algorithms}, pp. 343--362, 2012.

\bibitem{Bhosekarreview}
X.~Zhu and C.~Lai, ``Design and performance analysis of a magnetorheological
  fluid damper for drillstring,'' vol.~40, pp. 67--83, 01 2012.

\bibitem{poggio2017and}
T.~Poggio, H.~Mhaskar, L.~Rosasco, B.~Miranda, and Q.~Liao, ``Why and when can
  deep-but not shallow-networks avoid the curse of dimensionality: a review,''
  \emph{International Journal of Automation and Computing}, vol.~14, no.~5, pp.
  503--519, 2017.

\bibitem{alsaffar2018band}
Y.~Alsaffar, S.~Sassi, and A.~Baz, ``Band gap characteristics of nonrotating
  passive periodic drill string,'' \emph{Journal of Vibration and Acoustics},
  vol. 140, no.~2, p. 021004, 2018.

\bibitem{reviewghasemloonia}
H.~Moradi, F.~Bakhtiari-Nejad, and M.~Sadighi, ``Suppression of the bending
  vibration of drill strings via an adjustable vibration absorber,''
  \emph{International Journal of Acoustics and Vibration}, vol.~17, no.~3, p.
  155, 2012.

\bibitem{spanos1995modeling}
P.~Spanos, A.~Sengupta, R.~Cunningham, and P.~Paslay, ``Modeling of roller cone
  bit lift-off dynamics in rotary drilling,'' \emph{Journal of Energy Resources
  Technology}, vol. 117, no.~3, pp. 197--207, 1995.

\bibitem{jansen1995active}
J.~Jansen and L.~Van~den Steen, ``Active damping of self-excited torsional
  vibrations in oil well drillstrings,'' \emph{Journal of sound and vibration},
  vol. 179, no.~4, pp. 647--668, 1995.

\bibitem{chowdhury2013mixed}
S.~Chowdhury, W.~Tong, A.~Messac, and J.~Zhang, ``A mixed-discrete particle
  swarm optimization algorithm with explicit diversity-preservation,''
  \emph{Structural and Multidisciplinary Optimization}, vol.~47, no.~3, pp.
  367--388, 2013.

\bibitem{liu2012wave}
L.~Liu and M.~I. Hussein, ``Wave motion in periodic flexural beams and
  characterization of the transition between bragg scattering and local
  resonance,'' \emph{Journal of Applied Mechanics}, vol.~79, no.~1, 2012.

\bibitem{inversenets}
L.~Ardizzone, J.~Kruse, S.~Wirkert, D.~Rahner, E.~W. Pellegrini, R.~S. Klessen,
  L.~Maier-Hein, C.~Rother, and U.~K{\"o}the, ``Analyzing inverse problems with
  invertible neural networks,'' \emph{arXiv preprint arXiv:1808.04730}, 2018.

\bibitem{mccann2017convolutional}
M.~T. McCann, K.~H. Jin, and M.~Unser, ``Convolutional neural networks for
  inverse problems in imaging: A review,'' \emph{IEEE Signal Processing
  Magazine}, vol.~34, no.~6, pp. 85--95, 2017.

\end{thebibliography}

\end{document}